\documentclass[a4,reqno,10pt]{amsart}
\pdfoutput=1

\usepackage[foot]{amsaddr}

\usepackage{amsmath,amsthm,amssymb,xcolor,graphicx,soul}
\usepackage{amsfonts}
\usepackage{float}
\usepackage{amscd,verbatim,a4wide}
\usepackage[all]{xy}
\usepackage{braket}
\usepackage[mathscr]{euscript}
\usepackage{dsfont}
\usepackage{mathtools}
\usepackage[customcolors]{hf-tikz}
\usepackage{subcaption}
\usepackage{tikz}
\usepackage{tikz-cd}
\usepackage[most]{tcolorbox}
\usepackage{dsfont}

\usepackage{bbm}

\usepackage{hyperref}
\hypersetup{colorlinks=true,	citecolor = blue}  

\def\be{\begin{equation}}
\def\ee{\end{equation}}

\theoremstyle{plain}

\theoremstyle{definition}

\theoremstyle{remark}

\numberwithin{equation}{section}
\numberwithin{theorem}{section}
\numberwithin{figure}{section}
\numberwithin{table}{section}

\setstcolor{blue}

\usepackage{tikz}
\usetikzlibrary{arrows}
\usetikzlibrary{positioning}
\usepackage{amsmath}
\usepackage{amsfonts}
\usepackage{hyperref}
\usepackage{amsthm}
\usepackage{graphicx}

\newcommand{\Lie}{\mathcal{L}}

\newcommand{\fg}{\mathfrak{g}}
\newcommand{\GG}{\mathsf{G}}

\newcommand{\RR}{\mathbb{R}}

\newcommand{\pr}{\partial}
\DeclareMathOperator{\dd}{d\!}  
\newcommand{\ZZ}{\mathbb{Z}} 

\newcommand{\bT}{\mathbf{T}}
\newcommand{\bF}{\mathbf{f}}
\newcommand{\bQ}{\mathbf{Q}}
\newcommand{\bR}{\mathbf{R}}

\newcommand{\sg}{\mathsf{g}}
\newcommand{\od}{O(d,d)}

\newcommand*\Diff[1]{\mathop{}\!\mathrm{d^#1}}

\usepackage[english]{babel}
\usepackage{blindtext}

\begin{document}

\title[Non-abelian T-folds]{Non-abelian T-folds}

\author[M Bugden]{{Mark Bugden$^\dagger$} \\ \today}

\address{$^\dagger$Mathematical Sciences Institute, Australian National University, Canberra ACT 0200, Australia }

\email{mark.bugden@anu.edu.au}

\tikzset{node distance=1in, auto}

\maketitle

{
	\hypersetup{linkcolor=black}
\setcounter{tocdepth}{1}
	\tableofcontents
}

\begin{abstract}
We discuss the conditions under which non-abelian T-duality can be considered as a chain of abelian T-dualities. Motivated by these results, we propose that the topology of a non-abelian T-dual should be phrased in the language of T-folds, and give the explicit $\od$ transformations which can be used to glue the dual space. 
\end{abstract}


\section{Introduction}
Abelian T-duality, introduced by Buscher in \cite{Bus87,Bus88}, provides a prescription for relating two \emph{a priori} different string backgrounds. More specifically, it says that if we have a string background with an abelian isometry, then we can construct a new string background (which also has an abelian isometry) for which the physics is equivalent. Although primarily of interest in string theory, it is also useful from a purely supergravity perspective where it can be thought of as a solution generating technique. De la Ossa and Quevedo generalised the gauging procedure of Buscher to extend the technique to spaces admitting a non-abelian group of isometries \cite{DQ}, and this was later extended to include the Ramond-Ramond fields \cite{ST11, LOST}. Although the role that non-abelian T-duality plays in string theory is currently unclear, it has been employed successfully as a solution generating technique in supergravity \cite{AN,BGMNT,CMN,DMRV,INST,LORS,PRW,PTW,P,Z} and generalised supergravity \cite{HKO}. It has also been studied in the context of the $AdS$/CFT correspondence in \cite{GLMN,GZ,INNSZ,LMMN,LN,LNZ}.

One of the biggest open problems in T-duality is understanding the global nature of non-abelian T-duality. In other words, what is the topology of the T-dual space? Since the dual space is comprised of the base space $M$, together with the fibers described by the Lagrange multiplier coordinates, there are two components to this question: 
\begin{itemize}
	\item What is the topology of the fibers?
	\item How are these fibers patched together globally over the base?
\end{itemize}

For abelian T-duality, we have explicit answers to both of these questions. A string theoretic argument from higher-genus worldsheets says that the topology of the fibers is unchanged after a T-duality \cite{RV}. That is, if the original spacetime has circular fibers,\footnote{So that the Lagrange multiplier has a periodicity imposed on it.} then the dual spacetime will also have circular fibers. Similarly, if the fibers of the original space are non-compact, then the fibers of the dual space will also be non-compact. The analagous argument for higher-genus worldsheets is lacking for non-abelian T-duality, so we don't seem to have any information on the topology of the fibers (although there has been some work from an $AdS$/CFT perspective in, for example, \cite{LM,LOR,MNPRW} and related works).

The second question is often referred to as topological T-duality, and is also well-understood. For circle bundles, topological T-duality intermixes the topology of the total space with the $H$-flux \cite{BEM03,BEM04}. For torus bundles, T-duality can take us outside the realm of geometry/topology altogether and into non-commutative geometry/topology \cite{BHM03,BHM04,MR} (or even non-associative geometry! \cite{BHM05}). From a physics perspective, such non-commutative spaces are more often studied in the context of generalised geometry, where they are referred to as T-folds. We shall see that for non-abelian T-duality, the answer to the first question already involves T-folds. We do not attempt an answer to the second question.\footnote{For this reason, we will restrict ourselves in this paper to considering non-abelian T-duality for group manifolds. That is, we will ignore spectator coordinates. } \\

The simplest appearance of a T-fold occurs when considering a chain of T-dualities for a three-torus $\mathbb{T}^3$ with $H$-flux. The metric is simply the flat metric
\begin{align}
\dd s^2 &= \dd x^2 + \dd y^2 + \dd z^2,
\end{align}
and we wish to choose a $B$-field such that $H = \dd B$ is non-trivial in cohomology. Explicitly, we may take 
\begin{align}
B = -x \dd y \wedge \dd z,
\end{align}
so that 
\begin{align}
H = -\dd x \wedge \dd y \wedge \dd z. 
\end{align}
If we perform an abelian T-duality along $\pr_z$, we obtain a dual space known as the twisted torus:
\begin{subequations}
\label{eq:twistedtorus}
\begin{align}
\dd s^2 &= \dd x^2 + \dd y^2 + (\dd \hat{z} - x \dd y)^2 \\[1em]
B &= 0.
\end{align}
\end{subequations}
This space is a non-trivial circle bundle over $\mathbb{T}^2$. This dual model is known as the $f$-flux background, and provides us with clear example of how the gauge field and the geometry intermix under T-duality. A quick glance at \eqref{eq:twistedtorus} is enough to confirm that $\pr_y$ is still a Killing vector, and we can therefore perform another T-duality along it. The result is the so-called $Q$-flux background:
\begin{subequations}
\begin{align}
\dd s^2 &= \dd x^2 + \frac{1}{1+x^2} \left( \dd \hat{y}^2 + \dd  \hat{z}^2 \right) \\[1em]
B &= -\frac{x}{1+x^2} \dd \hat{y} \wedge \dd \hat{z} 
\end{align}
\end{subequations}
Note that a na\"{i}ve attempt to perform a third T-duality along $\pr_x$ fails, since $\pr_x$ is no longer an isometry of the metric. Despite this, the putative T-dual, known as the $R$-flux background, appears often in the literature particularly in the context of double field theory. We will not discuss this background.\\

The example of the torus with $H$-flux is oft-studied in the literature because it is quite simple, but nevertheless exhibits a lot of the interesting features of T-duality. The three torus directions of the original model provide, in principle, three isometries to gauge, and therefore three different T-dualities to perform. Indeed, the dual spaces we have obtained fit into a series of T-dualities: 
\begin{align}
\label{chainabelian}
\mathbf{T}_{xyz} \stackrel{\pr_z}{\longleftrightarrow} \mathbf{f}_{xy} \!^z \stackrel{\pr_y}{\longleftrightarrow} \mathbf{Q}_x \!^{yz} \stackrel{\pr_x}{\longleftrightarrow} \mathbf{R}^{xyz}.
\end{align}
In \cite{BBKW}, following the work of \cite{C16,CDJ15}, it was realised that there are some non-abelian T-dualities which give the same dual space as a chain of abelian T-dualities. The simplest example is the non-abelian T-dual of the $f$-flux background, which has a non-abelian Lie algebra of isometries. The chain of T-dualities is given by:
\begin{equation}
\label{fig:chainNATD}
\begin{tikzpicture}[baseline=(current  bounding  box.center)]
\node (T) {$\bT_{xyz}$};
\node (f1) [right=2cm of T] {$\bF^x \!_{yz}$};
\node (f2) [below=2cm of T] {$\bF_{xy}\!^z$};
\node (Q) [below=2cm of f1] {$\bQ^{xy}\!_z$};

\draw[<->] (T) to node {$\pr_x$} (f1);
\draw[<->] (f2) to node {$\pr_z$} (T);
\draw[->,dashed] (f2) to node [swap] {$NATD$} (Q);
\draw[<->] (f1) to node {$\pr_y$} (Q);
\draw[red,->] (T.135) arc (0:290:3mm) node[pos=0.5,swap]{$\delta B$};
\end{tikzpicture}
\end{equation}
where we must perform a gauge transformation for the $B$-field in order to continue the T-duality chain. In fact, there is an even nicer picture here. Starting with the three torus with flux, the various abelian T-dualities can be incorporated into the following duality cube:
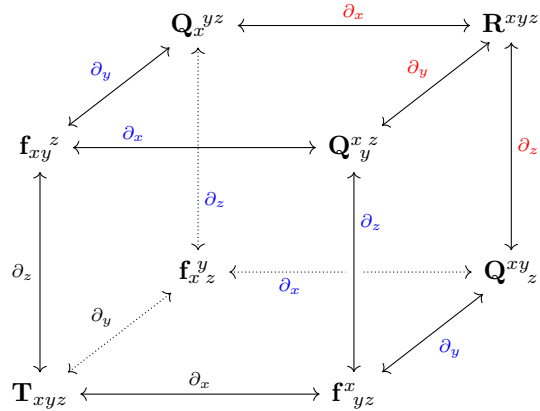
\begin{center}
	\begin{tikzcd}[back line/.style={densely dotted}, row sep=3em, column sep=3em]
		& \bQ_x\!^{yz} \ar[leftrightarrow]{rr}[red]{\pr_x} 
		& & \bR^{xyz}  \\
		\bF_{xy}\!^z \ar[leftrightarrow]{ur}[blue]{\pr_y} \ar[crossing over,leftrightarrow]{rr}[near start,blue]{\pr_x} 
		& & \bQ^{x}\!_y\!^z \ar[leftrightarrow]{ur}[red]{\pr_y} \\
		& \bF_{x}\!^y \!_z \ar[back line,leftrightarrow]{rr}[blue, near start,swap]{\pr_x}  \ar[back line,leftrightarrow]{uu}[near start,blue,swap]{\pr_z} 
		& & \bQ^{xy}\!_{z} \ar[leftrightarrow]{uu}[red,swap]{\pr_z} \\
		\bT_{xyz} \ar[leftrightarrow]{uu}{\pr_z} \ar[leftrightarrow]{rr}{\pr_x} \ar[back line,leftrightarrow]{ur}{\pr_y} & & \bF^x\!_{yz} \ar[crossing over,leftrightarrow]{uu}[blue, near end, swap]{\pr_z} \ar[leftrightarrow]{ur}[blue,swap]{\pr_y}
	\end{tikzcd}
\end{center}
The non-abelian T-duality we have been discussing then corresponds to a map from opposite sides of the cube, sending $f \to Q$. This can be realised as a chain of abelian T-dualities by tracing along the edges of the cube. One such path is the one we have already discussed in (\ref{fig:chainNATD}):
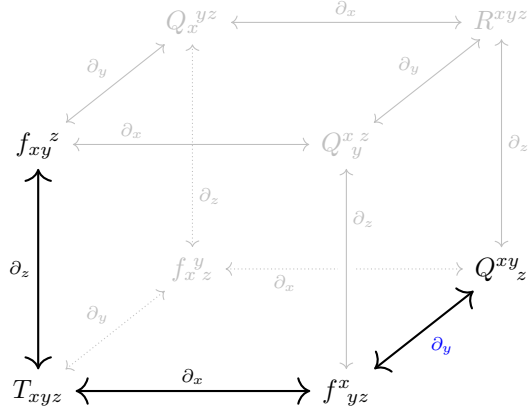
\begin{center}
	\begin{tikzcd}[back line/.style={densely dotted}, row sep=3em, column sep=3em]
		& \color{lightgray}{Q_x\!^{yz}} \ar[lightgray,leftrightarrow]{rr}[lightgray]{\pr_x} 
		& & \color{lightgray}{R^{xyz}}  \\
		f_{xy}\!^z \ar[lightgray,leftrightarrow]{ur}[lightgray]{\pr_y} \ar[crossing over,lightgray,leftrightarrow]{rr}[near start,lightgray]{\pr_x} 
		& & \color{lightgray}{Q^{x}\!_y\!^z} \ar[lightgray,leftrightarrow]{ur}[lightgray]{\pr_y} \\
		& \color{lightgray}{f_{x}\!^y \!_z} \ar[lightgray,back line,leftrightarrow]{rr}[lightgray, near start,swap]{\pr_x}  \ar[back line,lightgray,leftrightarrow]{uu}[near start,lightgray,swap]{\pr_z} 
		& & Q^{xy}\!_{z} \ar[lightgray,leftrightarrow]{uu}[lightgray,swap]{\pr_z} \\
		T_{xyz} \ar[leftrightarrow,thick]{uu}{\pr_z} \ar[thick,leftrightarrow]{rr}{\pr_x} \ar[back line,lightgray,leftrightarrow]{ur}{\pr_y} & & f^x\!_{yz} \ar[crossing over,lightgray,leftrightarrow]{uu}[swap,near end,lightgray]{\pr_z} \ar[thick,leftrightarrow]{ur}[blue,swap]{\pr_y}
	\end{tikzcd}
\end{center}
Note that the inverse map isn't well-defined as a single non-abelian T-duality, since the $Q$-flux background does not have a globally defined non-abelian group of isometries with which we can dualise. This is in line with our expectations - we aren't normally able to invert non-abelian T-duality. On the other hand, the inverse chain of abelian T-dualities is certainly well-defined. This coincidence - that a non-abelian T-duality agrees with a chain of abelian T-dualities - is more than a curiosity. Topological aspects of non-abelian T-duality are still not understood, even for the simplest examples of gauging the left action of a group on itself. The twisted torus, however, provides us with an example where we know \emph{explicitly} the topology of the non-abelian T-dual, since the non-abelian T-dual agrees with a chain of abelian T-dualities, whose topological behaviour is well-understood. In this case, the non-abelian T-dual is no longer a geometric space,\footnote{For a wonderful review on non-geometric backgrounds in string theory, see \cite{P18}.} so ``topological aspects of non-abelian T-duality'' should be suitably interpreted. The appearance of the $Q$-flux background as the non-abelian T-dual of the $f$-flux background suggests that global aspects of non-abelian T-duality might only be understood in the broader context of non-commutative and non-associative geometry and T-folds. Similar observations have been made in \cite{LO}.\\

The main purpose of this paper is twofold. First, we show that there is a class of non-abelian T-dualities that can be interpreted instead as a chain of abelian T-dualities. This observation sheds light on the topological nature of these non-abelian T-duals, since the topology of abelian T-duality is well-understood. The second purpose of this paper is to discuss, more generally, the relationship between arbitrary non-abelian T-duals and T-folds. 

Our paper is organised as follows: in Section \ref{sec:NATDchain} we discuss under which conditions we can view a non-abelian T-duality as a chain of abelian T-dualities. Then, in Section \ref{sec:NATfolds} we discuss the relationship between spaces obtained through non-abelian T-duality and T-folds. Finally, in Section \ref{sec:conclusion} we provide some concluding remarks.


\section{Non-abelian T-duality as a chain of abelian T-dualities}
\label{sec:NATDchain}


\subsection{Non-abelian T-duality and $O(d,d)$ transformations}

\begin{figure}
\begin{center}
	\begin{tikzpicture}[node distance = 3.5cm]
	\node (number1) {$S[X]$};
	\node (number2) [above right of = number1]{$S_G[X,A,\chi]$};
	\draw[->] (number1.east) .. controls   +(right:0.4cm) and +(down:0.4cm) .. node[below,rotate=50] {\small gauge isometries} (number2.south) ;

	\draw[->,blue] (number2.west) .. controls  +(left:0.4cm) and +(up:0.4cm)  .. node[above,rotate=55] {\small integrate $\chi$}  (number1.north) ;
	
	\draw[->,blue] (number2.west) .. controls  +(left:0.4cm) and +(up:0.4cm)  .. node[below,rotate=55] {\small and fix gauge }  (number1.north) ;

	\node (number3) [below right of = number2] {$\widetilde{S}[\chi]$};
	
	\draw[<-,red] (number3.north) .. controls  +(up:0.4cm) and +(right:0.4cm)  .. node[above,rotate=-55] {\small integrate $A$} (number2.east) ;
	
	\draw[<-,red] (number3.north) .. controls  +(up:0.4cm) and +(right:0.4cm)  .. node[below,rotate=-55] {\small and fix gauge} (number2.east) ;
	
	\end{tikzpicture}
\end{center}
\caption[]{A schematic of the Buscher procedure for T-duality.}
\label{fig:Buscher}
\end{figure}
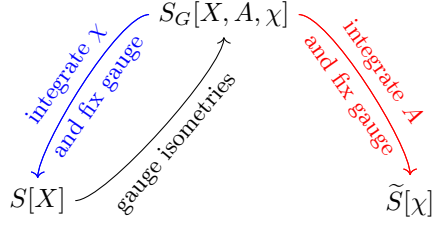

Let us review here the relevant features of non-abelian T-duality. We begin with a group manifold $\GG$ of dimension $d$. Let us choose a basis $\{T_a\}$ for the Lie algebra $\fg$ of $\GG$, and express the (left-invariant) Maurer--Cartan form $\lambda = \sg^{-1} \dd \sg$ in terms of this basis as $\lambda = \lambda^a T_a$. Consider now the $(0,2)$-tensor on $\GG$ given by 
\begin{align}
E &= E_{ij} \lambda^i \lambda^j,
\end{align}
where $E_{ij}$ is a constant, invertible matrix. Decomposing $E$ into the symmetric and antisymmetric parts gives us the metric and the $B$-field:
\begin{align}
E = g+B.
\end{align}
The non-linear sigma model corresponding to this background is
\begin{align}
S &= \int_{\Sigma} \Diff2z \, E_{ij} (\sg^{-1} \pr \sg)^i (\sg^{-1} \bar{\pr}\sg)^j.
\end{align}
Since the Maurer--Cartan forms are left-invariant, the right-invariant vector fields $R_a$ (which generate the left action) are symmetries of this action. That is, since 
\begin{align}
\Lie_{R_a} \lambda^b &= 0,
\end{align}
we have 
\begin{align}
\Lie_{R_a} E = 0.
\end{align}
To obtain the dual space, we follow the Buscher procedure shown schematically in Figure \ref{fig:Buscher}. First, one gauges the isometries by minimally coupling gauge fields $A$. One also introduces a Lagrange multiplier term $\chi$ which enforces a flat connection. Fixing a gauge then reduces back to the original model. On the other hand, integrating out the non-propagating gauge fields and fixing a gauge for the original coordinates gives us the T-dual model. 

The dual metric and $B$-field are given by
\begin{align}
\label{eq:NATDBuscher}
\widehat{E}_{ij} &:= NATD_{\GG} \left( E_{ij} \right)  = (E_{ij} + f^k_{ij} \chi_k)^{-1},
\end{align}
where $f^k_{ij}$ are the structure constants associated to the Lie algebra of right-invariant vector fields. 

We can see this non-abelian T-duality transformation as an $\od$ transformation, albeit a non-standard one,\footnote{Note that in Double Field Theory (DFT), the $O(d,d)$ transformation matrices are often taken to be constant, so that the transformed fields can still solve the strong constraint. Here, it is not at all clear that the transformed fields solve the strong constraint.} as follows. Recall that an $\od$ matrix
\begin{align}
\mathcal{O} &= \left( 
\begin{matrix}
a & b \\
c & d
\end{matrix}
\right)
\end{align}
acts on $E = g + B$ by fractional linear transformations:
\begin{align}
\mathcal{O} \cdot E &= (aE + b)(cE+d)^{-1}.
\end{align}
It is then easy to see that the non-abelian T-duality transformation \eqref{eq:NATDBuscher} is implemented by the following $\od$ matrix
\begin{align}
\label{eq:OddNATD}
\mathcal{O}_{NATD} &= \left( 
\begin{matrix}
0 & \mathds{1} \\
\mathds{1} & f^k_{ij} \chi_k
\end{matrix} \right) =
 \left( 
\begin{matrix}
0 & \mathds{1} \\
\mathds{1} & 0
\end{matrix} \right) \left( 
\begin{matrix}
\mathds{1} & f^k_{ij} \chi_k \\
0 & \mathds{1}
\end{matrix} \right) 
= \mathcal{O}_{\ZZ_2} \mathcal{O}_B
\end{align}
such that
\begin{align}
\widehat{E}  &= \mathcal{O}_{NATD} \cdot E.
\end{align}
That is, non-abelian T-duality can be thought of as first performing a $B$-shift by $f^k_{ij} \chi_k$, and then performing a factorised duality. Of course, this $B$-shift is not a gauge transformation of the $B$-field since it doesn't leave the field strength $H$-invariant. Indeed, this transformation involves the dual coordinates and so should more rightly be thought of as a transformation on the doubled space. For a discussion of non-abelian T-duality in the context of DFT, see \cite{Has}.  \\

The example of the $f$-flux background mentioned in the introduction shows that there is an example of a non-abelian T-duality where we can view it as a chain of abelian T-dualities, together with an honest $B$-field gauge transformation.\footnote{That is, without resorting to a doubled description.} In order to reconcile this with the decomposition \eqref{eq:OddNATD} we note that if the factorised duality, or factorised dualities, commute with $\mathcal{O}_B$ then the $B$-shift can be well-defined on the target space since the dual coordinates become coordinates after a factorised duality. This requirement places constraints on the Lie algebra, and in the next section we investigate the Lie algebras for which this can occur.


\subsection{Algebraic conditions}
Let us now discuss the conditions under which we can view non-abelian T-duality as a chain of abelian T-dualities. Let us assume that $\fg$, the Lie algebra of right-invariant vector fields, is a non-abelian Lie algebra of dimension $d$.\\

Consider the full factorised duality matrix $\mathcal{O}_{\ZZ_2}$. This can be written as a product of individual dualities:
\begin{align}
\label{eq:factorisedproduct}
\left( 
\begin{matrix}
0 & \mathds{1} \\
\mathds{1} & 0
\end{matrix} \right) &= \prod_{i = 1}^d \left( 
\begin{matrix}
\mathds{1} - E_i & E_i \\
E_i & \mathds{1} - E_i
\end{matrix} \right) = \prod_{i = 1}^d \mathcal{O}_i,
\end{align}
where $E_i$ is a $d \times d$ matrix with a $1$ in the $(i,i)$ entry and zeros elsewhere, and where the product in \eqref{eq:factorisedproduct} can be taken in any order since the $\mathcal{O}_i$ commute amongst themselves. In order for us to view non-abelian T-duality as a chain of abelian T-dualities, the $B$-shift has to involve only coordinates, and not dual coordinates. This means that if a dual coordinate appears explicitly in $\mathcal{O}_B$, then we need to perform a factorised duality in that coordinate \emph{before} we perform the $B$-shift. Said another way, if there is some $k$ for which $f^k_{ij}$ is non-zero, then we require that $\mathcal{O}_k$ commute with $\mathcal{O}_B$. Let us denote the set of these $k$ by $\Lambda$. That is, define
\begin{align}
\Lambda  &= \Big\{ k \in \{1,\dots, d\} : f^k_{ij} \not = 0\Big\}.
\end{align}
This splits the basis elements of $\fg$ according to whether they appear in the image of the Lie bracket or not, with $\{T_a\} = \{t_{\mu} , \widetilde{T}_{\nu}\}_{\mu \not \in \Lambda,\, \nu \in \Lambda}$. Then it follows that 
\begin{align}
\fg^1 := [\fg , \fg] \subset \textrm{span}  \{ \widetilde{T}_{\nu} \} .
\end{align}
The set of $\widetilde{T}_{\nu}$ form an ideal of $\fg$, which is non-zero since $\fg$ is not abelian. 

For $k \in \Lambda$, the commutator is schematically given by
\begin{align}
[\mathcal{O}_k, \mathcal{O}_B] &= \left( 
\begin{matrix}
-f \chi \cdot E & - [E,f\chi] \\
0 & f\chi \cdot  E
\end{matrix} \right).
\end{align} 
With indices written, it takes the form
\begin{align}
[\mathcal{O}_k, \mathcal{O}_B] &= \left( 
\begin{matrix}
-f^m_{ij} \chi_m E_k & - E_k f^m_{ij} \chi_m + f^m_{ij} \chi_m E_k \\
0 & -f^m_{ij} \chi_m E_k
\end{matrix} \right).
\end{align}
This vanishes precisely when $f^m_{ik} = 0$ for all $i$ and $m$. That is, $\mathcal{O}_k$ commutes with $\mathcal{O}_B$ when $[T_i, \widetilde{T}_k] = 0$ for all $i \in \{1,\dots, d\}$. It follows that $\textrm{span} \{\widetilde{T}_{\nu} \} \subset Z(\fg)$, where $Z(\fg)$ is the center of $\fg$. Thus in order for us to interpret a non-abelian T-duality as a chain of abelian T-dualities, we require the existence of a subalgebra $\mathfrak{h} = \textrm{span} \{\widetilde{T}_{\nu}\}$ such that
\begin{align}
\label{eq:algebracondition}
[\fg,\fg] \subset \mathfrak{h} \subset Z(\fg).
\end{align}
Note that $\mathfrak{h}$ is a non-trivial abelian ideal in $\fg$. We can see immediately that $\fg$ cannot be a semisimple Lie algebra, since that would imply that $\mathfrak{h} = Z(\fg) =  \fg$, which would imply that $\fg$ is abelian. In the special case that $[\fg,\fg] = \mathfrak{h} = Z(\fg)$, then $\fg$ is a 2-step nilpotent Lie algebra,\footnote{Recall that a nilpotent Lie algebra is $k$-step if $\fg^k:= [\fg^{k-1},\fg] = \{0\}$ and $\fg^{k-1} \not=\{0\}$.} and thus any 2-step nilpotent Lie algebra provides an example where we can consider non-abelian T-duality as a chain of abelian T-dualities. On the other hand, for $k$-step nilpotent algebras with $k \geq 3$ we see that $[\fg,\fg]$ is not abelian, and therefore cannot be contained in $Z(\fg)$. It follows that nilpotent Lie algebras only provide examples when they are 2-step.\\

It is worth mentioning that if $\mathfrak{h} = \textrm{span} \{ \widetilde{T}_{\nu} \}$ is an abelian ideal as required, then the $B$-shift defined by $\mathcal{O}_B$ is a shift by a closed form, and is therefore an honest gauge transformation of the $B$-field. That is, the $B$-shift leaves $H$-invariant.\footnote{Although the inversion may change $H$, as usual.}


\subsection{Examples: The Heisenberg groups}
\label{subsec:Heis}
The Heisenberg algebras provide a class of 2-step nilpotent Lie algebras, and therefore a class of examples in which we can view non-abelian T-duality as a chain of abelian T-dualities. Recall that the Heisenberg group $H_n$ of dimension $d = 2n+1$ is the set of $(n+2) \times (n+2)$ matrices of the form
\begin{align}
\left( 
\begin{matrix}
1 & \textbf{x} & z \\
0 & \mathds{1}_n & \textbf{y} \\
0 & 0 & 1
\end{matrix}
\right),
\end{align}
where $z$ is a real number, $\mathds{1}_n$ is the $n \times n$ identity matrix, $\textbf{x} = (x_1, \dots , x_n)$, and $y = (y_1 , \dots , y_n)^T$. The group structure is given by matrix multiplication. A convenient basis for the Lie algebra $\mathfrak{h}_n$ of $H_n$ is given by
\begin{subequations}
	\label{eq:HeisLAbasis}
\begin{align}
q_i &= 
\left( 
\begin{matrix}
0 & e_i & 0 \\
0 & 0 & 0 \\
0 & 0 & 0
\end{matrix}
\right) \\[1em]
p_j &= 
\left( 
\begin{matrix}
0 & 0 & 0 \\
0 & 0 & e_j^T \\
0 & 0 & 0
\end{matrix}
\right) \\[1em]
Z &= 
\left( 
\begin{matrix}
0 & 0 & 1 \\
0 & 0 & 0 \\
0 & 0 & 0
\end{matrix}
\right),
\end{align}
\end{subequations}
where $\{e_1,e_2,\dots,e_n\}$ is the canonical basis for $\RR^n$. The commutation relations for the Lie algebra are simply the canonical commutation relations:
\begin{subequations}
\begin{align}
[q_i , p_j] &= \delta_{ij} Z \\[1em]
[q_i, Z] &= 0 \\[1em]
[p_j,Z] &= 0.
\end{align}
\end{subequations}
The left-invariant Maurer--Cartan forms, $\lambda = \sg^{-1} \dd \sg$, expressed in this basis as $\lambda = \lambda^a T_a$ are:
\begin{subequations}
\begin{align}
\lambda^{q_i} &= \dd x_i \\[1em]
\lambda^{p_j} &= \dd y_j \\[1em]
\lambda^Z &= \dd z - \textbf{x} \cdot \dd \textbf{y}.
\end{align}
\end{subequations}
We now define a left-invariant metric by
\begin{align}
\dd s^2 &= \delta_{ij} \lambda^{q_i}\lambda^{q_j} + \delta_{ij} \lambda^{p_i} \lambda^{p_j} + \lambda^Z \lambda^Z \notag \\[1em]
\label{eq:HeisLIM}
&= \dd \textbf{x}^2 + \dd \textbf{y}^2 + (\dd z - \textbf{x} \cdot \dd \textbf{y})^2
\end{align}
and we take $B = 0$.\footnote{We have used the shorthand $\dd \textbf{x}^2 = \dd \textbf{x} \cdot \dd \textbf{x} = \dd x_1^2 + \dd x_2^2 + \dots + \dd x_n^2$.} This metric is left-invariant, so the right-invariant vector fields, $R_a$, are isometries. Their expressions are:
\begin{align}
R_{q_i} &= \pr_{x_i} + y_i \pr_z\\[1em]
R_{p_j} &= \pr_{y_j}\\[1em]
R_{Z} &= \pr_z.
\end{align}
We can perform a non-abelian T-duality on this background by gauging the isometric left action of the group on itself. The resultant T-dual is given by \eqref{eq:NATDBuscher}. It is:
\begin{align}
\widehat{E} &= \left( 
\begin{matrix}
\mathds{1} & -\hat{z} \mathds{1} & 0 \\
\hat{z}\mathds{1} & \mathds{1} & 0 \\
0 & 0 & 1
\end{matrix}
\right)^{-1} \notag \\[1em]
&= \left( 
\begin{matrix}
\frac{1}{1+\hat{z}^2} \mathds{1} & \frac{\hat{z}}{1+\hat{z}^2} \mathds{1} & 0 \\
\frac{-\hat{z}}{1+\hat{z}^2} \mathds{1} & \frac{1}{1+\hat{z}^2} \mathds{1} & 0 \\
0 & 0 & 1.
\end{matrix}
\right)
\end{align}
Extracting the symmetric and antisymmetric parts gives us the dual metric and the dual $B$-field:
\begin{subequations}
\label{eq:HeisNATD}
\begin{align}
\widehat{\dd s}^2 &= \frac{1}{1+\hat{z}^2} \left( \dd \hat{\textbf{x}}^2 + \dd \hat{\textbf{y}}^2 \right) + \dd \hat{z}^2 \\[1em]
\widehat{B} &= \frac{\hat{z}}{1+\hat{z}^2} \, \delta^{ij} \dd \hat{x}_i \wedge \dd \hat{y}_j
\end{align}
\end{subequations}

On the other hand, we can first perform a single abelian T-duality on \eqref{eq:HeisLIM} with respect to $\pr_z$. The result is the flat metric on the $(2n+1)$-torus $\mathbb{T}^{2n+1}$, together with a non-zero $B$-field:
\begin{subequations}
\begin{align}
\label{eq:TorusMetric}
\dd s^2 &= \dd \textbf{x}^2 + \dd \textbf{y}^2 + \dd \hat{z}^2 \\[1em]
\label{eq:TorusB}
B &= - \textbf{x} \cdot \dd \textbf{y} \wedge \dd \hat{z}.
\end{align}
\end{subequations}
We would like to perform additional abelian T-dualities on this background along the $2n$ Killing vectors $\{\pr_{x_1} ,\dots, \pr_{x_n}, \pr_{y_1},\dots , \pr_{y_n}\}$, however the explicit dependence of the $B$-field on the coordinates $\textbf{x}$ spoils our ability to do so. We can, however, perform the following (large) gauge transformation of the $B$-field:
\begin{align}
B \mapsto B' = B + \dd \, \Big( \hat{z}(\textbf{x} \cdot \dd \textbf{y}) \Big),
\end{align}
so that 
\begin{align}
\label{eq:TorusnewBfield}
B' &= - \hat{z} \, \delta^{ij} \dd x_i \wedge \dd y_j.
\end{align}
Note that the $H$-flux is given by 
\begin{align}
\label{eq:TorusH}
H &= - \delta^{ij} \dd x_i \wedge \dd y_j \wedge \dd \hat{z}.
\end{align}
The metric \eqref{eq:TorusMetric} and $B$-field \eqref{eq:TorusnewBfield} now satisfy
\begin{align*}
\Lie_{\pr_{x_i}} g &= \Lie_{\pr_{x_i}} B = 0 \\
\Lie_{\pr_{y_j}} g &= \Lie_{\pr_{y_j}} B = 0,
\end{align*}
and so we can proceed to perform multiple abelian T-dualities along the vector fields $\{\pr_{x_1} ,\dots, \pr_{x_n}, \pr_{y_1},\dots , \pr_{y_n}\}$.
The formula for multiple abelian T-dualities is relatively easy to derive (see, for example, \cite{MBPhD}). If the coordinates $X^M$ of the manifold split into coordinates on the fibers $X^{m}$, and coordinates along the base $X^{\mu}$, then the metric and $B$-field decompose as
\begin{align}
E_{MN}&= 
\left(
\begin{matrix}
E_{mn} & E_{m \nu} \\
E_{\mu n} & E_{\mu \nu}
\end{matrix}
\right),
\end{align}
and the T-dual for the multiple abelian T-dualities is given by
\begin{align}
\label{eq:multipleBuscher}
\widehat{E}_{MN} &= 
\left(
\begin{matrix}
\widehat{E}_{mn} & \widehat{E}_{m \nu} \\
\widehat{E}_{\mu n} & \widehat{E}_{\mu \nu}
\end{matrix}
\right) = 
\left(
\begin{matrix}
(E^{-1})_{mn} & -(E^{-1})_{mn} E_{n \nu} \\
E_{\mu m} (E^{-1})_{mn} & E_{\mu \nu} - E_{\mu m} (E^{-1})_{mn} E_{n \nu}
\end{matrix}
\right).
\end{align}
In our case, we have 
\begin{subequations}
\begin{align}
E_{mn} &= 
\left(
\begin{matrix}
\mathds{1} & -\hat{z} \mathds{1} \\
\hat{z} \mathds{1} & \mathds{1} 
\end{matrix}
\right) \\[1em]
E_{\mu n} &= \left( \begin{matrix}
0 & \dots & 0
\end{matrix}\right) \\
E_{m \nu} &= \left(
\begin{matrix}
0 \\
\vdots \\
0
\end{matrix}
\right) \\
E_{\mu \nu} &= \left(
\begin{matrix}
1
\end{matrix}
\right).
\end{align}
\end{subequations}
The dual metric and $B$-field obtained from the $2n$ abelian isometries via \eqref{eq:multipleBuscher} are given by:
\begin{subequations}
\label{eq:HeisChain}
\begin{align}
\dd s^2 &= \frac{1}{1+\hat{z}^2} \left( \dd \hat{\textbf{x}}^2 + \dd \hat{\textbf{y}}^2 \right) + \dd \hat{z}^2 \\[1em]
B &= \frac{\hat{z}}{1+\hat{z}^2} \, \delta^{ij} \dd \hat{x}_i \wedge \dd \hat{y}_j ,
\end{align}
\end{subequations}
which is precisely the background \eqref{eq:HeisNATD} obtained from the non-abelian T-duality.
Note that when $n=1$ we get the situation described in the introduction and shown in \eqref{fig:chainNATD}.


\subsection*{A comment on the topology}

The (left-invariant) metric is given by
\begin{align}
\label{eq:Heismetric}
\dd s^2 &= \dd \textbf{x}^2 + \dd \textbf{y}^2 + (\dd z - \textbf{x} \cdot \dd \textbf{y})^2.
\end{align}
This is na\"{i}vely a curved metric on $\RR^{2n+1}$,\footnote{The metric has a scalar curvature of $-1$.} however, we are interested in a compact version of this. The metric \eqref{eq:Heismetric} is manifestly invariant under the identifications $y_i \sim y_i + 1$, and $z \sim z+1$. Under the identifications $x_i \sim x_i + 1$ we see that the metric is only invariant provided that $z$ is identified as $z \sim z+ y_i$. Written more succinctly, we have
\begin{align*}
(\textbf{x},\textbf{y},z) &\sim (\textbf{x}, \textbf{y},z+1) \sim (\textbf{x}, \textbf{y} + e_i, z) \sim (\textbf{x} + e_i, \textbf{y},z + y_i) 
\end{align*}
It follows that we can view this compact manifold as a (non-trivial) circle bundle over $\mathbb{T}^{2n}$.\footnote{The fibers of the bundle are parameterised by $z$, and the base $\mathbb{T}^{2n}$ is paramtererised by $(\textbf{x}, \textbf{y})$.} Indeed, circle bundles over $\mathbb{T}^{2n}$ are classified (up to bundle isomorphism) by elements of $H^2(\mathbb{T}^{2n},\ZZ)$. The class determining the isomorphism class of the bundle is the first Chern class of the associated line bundle. We can obtain an explicit representative of this class by computing the curvature $F = \dd A$ of a principal $U(1)$-connection. A glance at \eqref{eq:Heismetric} is enough to confirm that $A =\dd z - \textbf{x} \cdot \dd \textbf{y}$ is a principal $U(1)$-connection for the bundle, and that therefore the isomorphism class of the bundle is determined by the (non-trivial) class 
\begin{align}
\label{eq:HeisF}
F &= - \delta^{ij} \dd x_i \wedge \dd y_j.
\end{align}
Performing a T-duality along the circle fiber parameterised by $z$ gives a dual circle bundle which is determined up to bundle isomorphism by its first Chern class. From \cite{BEM03}, the first Chern class of the T-dual circle bundle is given by
\begin{align}
\widehat{F} &= \int_{S^1} H = 0,
\end{align}
so that the dual bundle is just the trivial circle bundle over $\mathbb{T}^{2n}$, that is, $\mathbb{T}^{2n+1}$. According to \cite{BEM03}, the dual flux should satisfy
\begin{align}
\int_{\widehat{S}^1} \widehat{H} &= F,
\end{align}
and we can see that this is satisfied by comparing \eqref{eq:TorusH} and \eqref{eq:HeisF}.

When we perform the additional $2n$ abelian T-dualities on $\mathbb{T}^{2n+1}$ with flux, the topological description of \cite{BEM03} is no longer sufficient, and we need to use the generalisation for torus bundles in \cite{MR,BHM05}. Here, we notice that since the $H$-flux has two `legs' along the toroidal fibers we are T-dualising along, the dual space will no longer be a principle torus bundle. Following \cite{MR,BHM05}, we could interpret the dual space as a bundle of non-commutative tori over the base $S^1$. Instead, we will take the perspective of T-folds.


\subsection{A non-example: The $4 \times 4$ unipotent upper triangular matrices}
Before we move on to Section \ref{sec:NATfolds} to discuss T-folds in the context of non-abelian T-duality, let us include a non-example which is a 3-step nilpotent Lie algebra. Consider the Lie group of $4 \times 4$ unipotent upper triangular matrices. That is, the set of matrices of the form
\begin{align}
\left( 
\begin{matrix}
1 & x_1 & x_2  & z \\
0 & 1 & w & y_1  \\
0 & 0 & 1 & y_2 \\
0 & 0 & 0 & 1
\end{matrix}
\right),
\end{align}
where $(x_1,x_2,y_1,y_2,w,z) \in \RR^6$. A convenient basis for the Lie algebra is given by the same basis for the Heisenberg algebra $H_2$, \eqref{eq:HeisLAbasis}, together with an additional element $W$:
\begin{align}
W &= 
\left(
\begin{matrix}
0 & 0 & 0 & 0 \\
0 & 0 & 1 & 0  \\
0 & 0 & 0 & 0 \\
0 & 0 & 0 & 0
\end{matrix}
\right).
\end{align}
The non-zero commutation relations are
\begin{subequations}
\begin{align}
[q_i , p_j] &= Z \\
[q_1,W] &= q_2 \\
[p_2,W] &= -p_1,
\end{align}
\end{subequations}
and the left-invariant Maurer--Cartan forms with respect to this basis are
\begin{subequations}
\begin{align}
\lambda^{q_1} &= \dd x_1 \\
\lambda^{q_2} &= \dd x_2 - x_1 \dd w \\
\lambda^{p_1} &= \dd y_1 - w \dd y_2 \\ 
\lambda^{p_2} &= \dd y_2 \\
\lambda^{W} &= \dd w \\
\lambda^{Z} &= \dd z +(w x_1 - x_2) \dd y_2 - x_1 \dd y_1.
\end{align}
\end{subequations}
Defining a left-invariant metric by
\begin{align}
\dd s^2 &= \delta_{ij} \lambda^{q_i}\lambda^{q_j} + \delta_{ij} \lambda^{p_i} \lambda^{p_j} + \lambda^W \lambda^W  +\lambda^Z \lambda^Z
\end{align}
gives a metric of constant scalar curvature $\mathcal{R} = -2$. The right-invariant vector fields are a non-abelian algebra of isometries, and we can gauge these isometries to construct a non-abelian T-dual. Note, however, that we cannot consider this non-abelian T-duality as a chain of abelian T-dualities, since
\begin{align}
[\fg ,\fg] &= \textrm{span} \{ q_2,p_1,Z \}
\end{align}
and 
\begin{align}
Z(\fg) &= \textrm{span} \{Z\},
\end{align}
from which it follows that there is no ideal $\mathfrak{h}$ such that \eqref{eq:algebracondition} holds.


\section{Non-abelian T-duals as T-folds}
\label{sec:NATfolds}
The dual space \eqref{eq:HeisChain} obtained as the chain of abelian T-dualities (or, equivalently, as a single non-abelian T-duality \eqref{eq:HeisNATD}) is not a globally-defined Riemannian manifold. Although we have a perfectly well-defined local description, it is easy to see that the metric and $B$-field are not globally-defined since the transformation $\hat{z} \sim \hat{z}+1$ does not leave them invariant. That is, there is no diffeomorphism relating $E(\textbf{x}, \textbf{y}, \hat{z})$ and $E(\textbf{x},\textbf{y},\hat{z}+1)$. On the other hand, if we allow ourselves a broader class of gluing functions, then it is possible to make sense of this background globally. A T-fold is a space in which the gluing functions are allowed to take values in $O(d,d)$, rather than just diffeomorphisms \cite{Hull04,Hull06}. A simple calculation shows that for $E$ defined by \eqref{eq:HeisChain}, we have
\begin{align}
E(\textbf{x},\textbf{y},z+1) &= \mathcal{O}_{\beta} \cdot E(\textbf{x},\textbf{y},z),
\end{align}
where $\mathcal{O}_{\beta}$ is the following $\beta$ transformation:
\begin{align}
\mathcal{O}_{\beta} &= 
\left( 
\begin{matrix}
\mathds{1}_{2n+1} & 0 \\
\beta & \mathds{1}_{2n+1}
\end{matrix}
\right),
\end{align}
and 
\begin{align}
\beta &= \left( 
\begin{matrix}
0 & \mathds{1}_n & 0\\
-\mathds{1}_n & 0 & 0 \\
0 & 0 & 0
\end{matrix}
\right)
\end{align}
Indeed, we can expand this description much more generally to include all non-abelian T-duals! Recall from \eqref{eq:NATDBuscher} that the non-abelian T-dual model $\widehat{E}$ can be expressed as an $O(d,d)$ transformation of the original model $E$ via
\begin{align}
\widehat{E}_{ij} &= (E_{ij} + f^k_{ij} \chi_k )^{-1} = \mathcal{O}_{\ZZ_2} \cdot \mathcal{O}_{B} \cdot E.
\end{align}
The dual model depends only on the dual coordinates $(\chi_1,\dots, \chi_d)$, and it turns out that it is possible to glue these models using $O(d,d)$ transformations. To see this, consider increasing one of the dual coordinates by 1. Then
\begin{align*}
\widehat{E}_{ij} (\chi_m + 1) &= (E_{ij} + f^k_{ij}\chi_k + f^k_{ij} \delta_{km})^{-1} \\
&= \mathcal{O}_{\ZZ_2} \cdot (E_{ij} + f^k_{ij}\chi_k + f^k_{ij} \delta_{km}) \\
&= \mathcal{O}_{\ZZ_2} \cdot \mathcal{O}_B \cdot (E_{ij} + f^k_{ij}\chi_k) \\
&= \mathcal{O}_{\ZZ_2} \cdot \mathcal{O}_B \cdot \mathcal{O}_{\ZZ_2} \cdot (E_{ij} + f^k_{ij}\chi_k)^{-1} \\
&= \mathcal{O}_{\ZZ_2} \cdot \mathcal{O}_B \cdot \mathcal{O}_{\ZZ_2} \cdot \widehat{E}_{ij} \\
&= \mathcal{O}_{\beta} \cdot \widehat{E}_{ij},
\end{align*}
where $\mathcal{O}_{B}$ is a $B$-shift by $f^k_{ij} \delta_{km}$. More generally, the value of $\widehat{E}$ at the point $\chi = (\chi_1 , \dots, \chi_d)$ is related to the value of $\widehat{E}$ at the point $\chi + \chi' = (\chi_1 + \chi_1', \dots , \chi_d + \chi_d')$ via
\begin{align}
\widehat{E}_{ij}(\chi + \chi') &= \mathcal{O}_{\beta} \cdot \widehat{E}_{ij} (\chi),
\end{align}
where
\begin{align}
\mathcal{O}_{\beta} &= 
\left( 
\begin{matrix}
\mathds{1} & 0 \\
f^k_{ij} \chi_k' & \mathds{1}
\end{matrix}
\right)
\end{align}


\subsection{Example: The NATD of $S^3$}
By far the most prevalent example of non-abelian T-duality in the literature is the non-abelian T-duality of the round three sphere, thought of as the group manifold $SU(2)$. We will take $B=0$ for simplicity. The round metric is the (unique) bi-invariant metric, and so is invariant under the left action of $SU(2)$, as well as the right action of $SU(2)$. In fact, the full isometry group is $SO(4) = \left( SU(2) \times SU(2) \right) / \ZZ_2$. Performing a non-abelian T-duality with respect to one of these isometries gives the following dual space, defined on the Lagrange multiplier coordinates $\{\chi_i\} = \{x,y,z\}$:
\begin{align}
\widehat{E}(x,y,z) &= \frac{1}{1+x^2+y^2+z^2} 
\left( 
\begin{matrix}
1+x^2 & xy - z & xz + y \\
xy + z & 1+y^2 & yz - x \\
xz - y & yz + x & 1+z^2
\end{matrix}
\right).
\end{align}
Extracting the symmetric and antisymmetric parts gives us the metric and $B$-field:
\begin{subequations}
\begin{align}
\widehat{\dd s}^2 &= \frac{1}{1+\chi^2} \left(\delta_{ij} + \chi_i \chi_j \right) \dd \chi^i \dd \chi^j \\[1em]
\widehat{B} &= -\epsilon_{ijk} \frac{\chi_k}{1+\chi^2} \dd \chi^i \wedge \dd \chi^j,
\end{align}
\end{subequations}
where $\chi^2 = \chi_1^2 + \chi_2^2 + \chi_3^2.$ Note that we can't view this non-abelian dual as a chain of abelian T-duals \`{a} la Section \ref{sec:NATDchain}, since $\mathfrak{su}(2)$ is a simple Lie algebra, and therefore has no non-trivial ideals.

Like all T-dualities, this metric is \emph{a priori} defined on the dual Lie algebra $\mathfrak{su}(2)^{\ast}$, which, since it is a vector space, is topologically $\RR^3$. The dual data takes a simpler form upon switching to polar coordinates via
\begin{subequations}
\begin{align}
x &= r \sin \theta \cos \phi \\[1em]
y &= r \sin \theta \sin \phi \\[1em]
z &= r \cos \theta,
\end{align}
\end{subequations}
from which we get the expressions
\begin{subequations}
\begin{align}
\widehat{\dd s}^2 &= \dd r^2 + \frac{r^2}{1+r^2} \left( \dd \theta^2 + \sin^2 \theta \dd \phi^2 \right) \\[1em]
\widehat{B} &= - \frac{r^3}{1+r^2} \sin \theta \dd \theta \wedge \dd \phi.
\end{align}
\end{subequations}
This geometry interpolates between $\RR^3$ near $r = 0$ and $\RR \times S^2$ as $r \to \infty$. The Ricci scalar is 
\begin{align}
\mathcal{R} &= \frac{2(r^4 + 3r^2 + 9)}{(1+r^2)^2},
\end{align}
from which it is clear that there is no radial compactification which results in a smooth geometry. In other words, there is no diffeomorphism relating $\widehat{E}(r,\theta,\phi)$ and $\widehat{E}(r+r',\theta,\phi)$. On the other hand, a short calculation shows that
\begin{align}
\widehat{E}(x+x',y+y',z+z') &= \mathcal{O}_{\beta} \cdot \widehat{E}(x,y,z),
\end{align}
where 
\begin{align}
\mathcal{O}_{\beta} &= 
\left( 
\begin{matrix}
\mathds{1} & 0 \\
\beta  & \mathds{1}
\end{matrix}
\right)
\end{align}
and 
\begin{align}
\beta &= \left( 
\begin{matrix}
0 & z' & -y' \\
z' & 0 & x' \\
y' & -x' & 0
\end{matrix}
\right).
\end{align}
It follows that we can view the non-abelian T-dual of $S^3$ as a compact space provided that we are allowed to glue using $O(d,d)$ transformations.


\section{Summary and Outlook}
\label{sec:conclusion}

In the first part of the paper we have discussed under which conditions we can consider non-abelian T-duality to be equivalent to performing a chain of abelian T-dualities. We found that the algebra of isometries, $\fg$, was required to have a non-trivial ideal $\mathfrak{h}$ such that 
\begin{align}
[\fg,\fg] \subset \mathfrak{h} \subset Z(\fg),
\end{align}
where $Z(g)$ is the center of the Lie algebra. This condition highlights the class of 2-step nilpotent Lie algebras as candidates, whilst excluding all simple/semisimple Lie algebras and $k$-step nilpotent Lie algebras for $k \geq 3$. We then discussed a class of examples (the Heisenberg algebras) generalising the well-known $f$-flux background from string theory, and showed that T-folds/non-commutative spaces appear naturally in the context of non-abelian T-duality.\\

The second part of the paper involved exploring the relationship between non-abelian T-duality and T-folds. The non-abelian Buscher rules are written as an $O(d,d)$ transformation of $E = g+B$, and this is used to show that incrementing a coordinate in the non-abelian T-dual can be similarly implemented with an $O(d,d)$ transformation. Thus the dual space can be considered as a compact manifold, with the metric and $B$-field at different points related by means of $O(d,d)$ transformations. Such transformations do not lie in the geometric subgroup, and therefore the dual space is not a Riemannian manifold but is instead a T-fold. Note that a generalisation to Poisson--Lie T-duality for the content of Section \ref{sec:NATfolds} follows relatively straightforwardly. We have not included this minor generalisation for ease of exposition. \\

An obvious and difficult question is how we should make the identifications in the dual space. For the $Q$-flux background obtained as the non-abelian T-dual of the $f$-flux background in \eqref{chainabelian} (or the generalisation in Section \ref{subsec:Heis}), this has an easy answer since we can simply compare to the space obtained as a chain of abelian T-dualities. In that case, we take the coordinate $z$ to be periodic, with periodicity determined by the genus-one worldsheet argument as in \cite{RV}. An alternative way to determine the periodicity of the $z$ coordinate is to consider bounds on the quantity 
\begin{align}
b &= \int_{\Sigma_2} B,
\end{align}
where $\Sigma_2$ is a suitably chosen closed 2-manifold, as in \cite{LM,LOR}. String theory arguments require this quantity to be bounded in $[0,1]$, and imposing this bound on the non-abelian T-dual of the $f$-flux gives us the correct periodicity for the $z$-coordinate. This argument has been used for the non-abelian T-dual of $S^3$ in \cite{LM,LOR}, where it was suggested that the radial coordinate be constrained to an interval $[n,n+1]$. The results of this paper suggest that, rather than having a large number of such intervals, one should glue the endpoints of the interval using $O(d,d)$ transformations. It would be interesting to further investigate the role of such transformations in $AdS$/CFT, and in particular to clarify the nature of this gluing procedure in the associated CFT.


\section*{Acknowledgements}
I would like to thank the organisers and participants of the workshop ``Dualities and Generalized Geometries" 9-16 September 2018, Corfu, Greece, and the organisers and participants of the workshop ``String and M-theory: The New Geometry of the 21st Century" 10-14 December 2018, Singapore, for the inspiring atmospheres present at both workshops. I would also like to thank Kyle Wright for his comments on a draft version of this paper, and David Berman for helpful discussions. 
This research was partially supported by the Australian Government through the Australian Research Council's Discovery Projects funding scheme (projects DP150100008 and DP160101520). 

      

\end{document}